\documentclass[sigconf]{acmart}
\usepackage{xspace}
\usepackage{graphicx}
\usepackage{hyperref}
\usepackage{booktabs}
\usepackage{array}
\usepackage{multirow}
\usepackage{multicol}
\usepackage{amsmath}
\usepackage{stfloats}
\newcommand{\quoted}[1]{\textit{``#1''}\xspace}

\AtBeginDocument{%
  }

\setcopyright{acmlicensed}
\copyrightyear{2025}
\acmYear{2025}
\setcopyright{cc}
\setcctype{by}
\acmConference[FAccT '25]{The 2025 ACM Conference on Fairness,
Accountability, and Transparency}{June 23--26, 2025}{Athens, Greece}
\acmBooktitle{The 2025 ACM Conference on Fairness, Accountability, and
Transparency (FAccT '25), June 23--26, 2025, Athens,
Greece}\acmDOI{10.1145/3715275.3732090}
\acmISBN{979-8-4007-1482-5/2025/06}

\begin{document}

\title{A Day in Their Shoes: Using LLM-Based Perspective-Taking Interactive Fiction to Reduce Stigma Toward Dirty Work}

\renewcommand{\shorttitle}{A Day in Their Shoes}

\author{Xiangzhe Yuan}
\orcid{0009-0001-2457-7985}
\affiliation{\institution{Department of Computer Science \\ University of Iowa}
\state{Iowa}
\country{USA}}

\author{Jiajun Wang}
\orcid{0009-0004-2584-889X}
\affiliation{\institution{School of Information Systems and Technology Management \\ University of New South Wales}
\city{Sydney}
\country{Australia}}

\author{Qian Wan}
\orcid{0000-0002-4250-8780}
\affiliation{\institution{Department of Computer Science \\ City University of Hong Kong}
\country{Hong Kong SAR}}

\author{Siying Hu}
\authornote{Corresponding author.}
\orcid{0000-0002-3824-2801}
\affiliation{\institution{Department of Computer Science \\ City University of Hong Kong}
\country{Hong Kong SAR}}

\renewcommand{\shortauthors}{Yuan et al.}

\begin{abstract}
Occupations referred to as ``dirty work'' often face entrenched social stigma, which adversely affects the mental health of workers in these fields and impedes occupational equity. In this study, we propose a novel Interactive Fiction (IF) framework powered by Large Language Models (LLMs) to encourage perspective-taking and reduce biases against these stigmatized yet essential roles. Through an experiment with participants (n = 100) across four such occupations, we observed a significant increase in participants' understanding of these occupations, as well as a high level of empathy and a strong sense of connection to individuals in these roles. Additionally, qualitative interviews with participants (n = 15) revealed that the LLM-based perspective-taking IF enhanced immersion, deepened emotional resonance and empathy toward ``dirty work,'' and allowed participants to experience a sense of professional fulfillment in these occupations. However, participants also highlighted ongoing challenges, such as limited contextual details generated by the LLM and the unintentional reinforcement of existing stereotypes. Overall, our findings underscore that an LLM-based perspective-taking IF framework offers a promising and scalable strategy for mitigating stigma and promoting social equity in marginalized professions.
\end{abstract}



\maketitle

\section{Introduction}
``Dirty work'' refers to occupations that face social stigma due to their unpleasant, hazardous, or morally questionable nature~\cite{ashforth1999can,ashforth2014dirty,simpson2012introducing,e2007normalizing}. Workers in these fields play an essential role in society, yet they often endure severe social stigma~\cite{yan2017dirty,osuna2017being,bosmans2016dirty,dick2005dirty,soni2013dirty}. Existing research shows that the stigma associated with dirty work not only triggers identity crises and psychological stress for these workers -- resulting in various mental and physical health issues -- but also contributes to organizational inefficiency~\cite{yan2017dirty,lemmergaard2012golfing}. While there is substantial research on dirty work itself, the exploration of methods to reduce public bias against these occupations is limited~\cite{ashforth2014dirty,yan2017dirty}.

Perspective-taking~\cite{underwood1982perspective,flavell2013perspectives} has been extensively studied as a way to reduce stigma and weaken negative stereotypes by fostering empathy towards specific social targets~\cite{chung2013reducing,tompkins2015reducing}. And empathy has been shown to reduce stigma by improving attitudes toward stigmatized groups as a whole ~\cite{batson1997empathy}. With the development of large language models (LLMs), a new avenue has emerged for facilitating perspective-taking in a more effective and scalable manner. LLMs have demonstrated the potential to generate personalized and context-rich scenarios~\cite{wang2024survey}, allowing users to immerse themselves in simulated real-life situations~\cite{li2024more}. Unlike traditional perspective-taking methods that rely on predefined scripts, LLMs can dynamically adapt scenarios based on users' responses, providing a flexible and personalized experience.

In this study, we further harness the capabilities of LLM within an Interactive Fiction (IF) framework. IF is a text-based narrative form that encourages users to make meaningful choices, thereby experiencing branching storylines~\cite{montfort2005twisty,niesz1984interactive}. By leveraging an LLM to generate and adapt these narratives, our approach offers greater immediacy and flexibility compared to traditional IF~\cite{basavatia2023complexworld}; potentially allowing for deeper engagement in immersive scenarios. By engaging participants in this experience, we seek to broaden the understanding of occupational realities and advance social equity.

This study focuses on two categories of dirty work: physical dirty work and social dirty work—further divided into four representative occupations: \textbf{Janitors} (engaging in physically unpleasant tasks), \textbf{Firefighters} (operating in hazardous environments), \textbf{Police officers} (frequently interacting with marginalized groups), and \textbf{Caregivers} (performing low-status, often menial tasks)~\cite{ashforth1999can,yan2017dirty}. Using LLM-based perspective-taking IF to generate realistic and personalized work scenarios, we try to address the following key research questions:

\textbf{RQ1:} To what extent does an LLM-based perspective-taking IF reduce social bias and stigma toward dirty workers?

\textbf{RQ2:} How might LLM-based perspective-taking influence perceptions of occupational stigma and social bias?

\textbf{RQ3:} What challenges arise in using LLM-based perspective-taking IF to reduce occupational stigma and social bias?

To investigate these research questions, we recruited 100 participants and randomly assigned them to one of four groups (n = 25 per group), each representing a distinct type of dirty work (janitors, firefighters, police officers, and caregivers). Before participating in the LLM-based perspective-taking IF experience, participants completed a pre-test that assessed empathy-related traits (e.g., Interpersonal Reactivity Index, beliefs about empathy) and their baseline knowledge of the assigned profession. Following the experiment, participants engaged in the perspective-taking IF customized by GPT-4o and then completed post-test assessments. Our quantitative analysis revealed a significant increase in participants’ understanding of these professions and elevated levels of empathy. Additionally, qualitative interviews (n = 15) indicated that, despite certain challenges, the immersive nature of LLM-based IF enhanced participants’ emotional resonance, increased empathy, and fostered greater respect for dirty workers. In summary, these findings suggest that LLM-based perspective-taking IF has the potential to serve as a novel and effective tool for reducing stigma and promoting a more equitable social attitude toward dirty work.
This research makes the following contributions:
\begin{itemize}
 \item This paper introduces LLM-based perspective-taking IF as a new approach to simulating immersive and personalized dirty work experiences. This provides the public with an innovative way to understand dirty work and reduce stigma.

 \item Through empirical research, we evaluate the effectiveness of this method in mitigating occupational stigma and offer insights into the challenges and opportunities of using AI to reduce bias. This study provides a new perspective on the application of LLMs in promoting social equity.
\end{itemize}

\section{Related work}
We review the concept and stigma associated with dirty work, highlighting perspective-taking as a method for stigma reduction. Additionally, we briefly discuss the capacity of large language models in interactive fiction.

\subsection{Concepts and Stigma of Dirty Work}
The concept of dirty work is used to describe occupations that are devalued, despised, or disliked by society. This term refers to jobs characterized by repulsiveness, unpleasantness, danger, low status, immorality, or disdain~\cite{yan2017dirty}. Hughes et al.~\cite{hughes2017sociological} argued that while ``dirty work'' serves essential social functions, it faces marginalization, which leads to the stigmatization of these occupations. As a result, such jobs bear a certain social stigma. Ashforth et al.~\cite{ashforth1999can} contended that the negative reactions elicited by ``dirty work'' arise from the violation of social standards for ``normal'' or ``clean'' work. The public attaches negative labels to these jobs, fostering prejudice and discrimination, which in turn leads to the stigmatization of workers in these roles. Consequently, ``dirty work`` is intrinsically linked to the stigmatization of occupations. The term ``dirty'' does not inherently describe the work itself nor the personal characteristics of the workers, but rather reflects social and individual perceptions ~\cite{simpson2012introducing}.

Ashforth et al.~\cite{ashforth1999can} classified dirty work into three categories based on its characteristics: physical dirty work, social dirty work, and moral dirty work. Physical dirty work is further divided into two types: jobs that are directly associated with unpleasant or dirty physical entities, or jobs conducted in environments considered harmful or dangerous~\cite{snyder1986distancing,ashforth2014dirty}. Social dirty work can be grouped into two main categories: work that entails frequent contact with marginalized or stigmatized groups and work that involves low-status, often menial tasks rooted in servitude. In both forms, just as individuals may be tainted by others’ glory, they can also be tainted by the perceived flaws of those they serve or interact with, ultimately finding themselves in positions deemed inferior relative to others ~\cite{snyder1986distancing,ashforth2014dirty}. Moral dirty work is viewed as generating social harm or violating social virtues, often involving deception or breaching social etiquette standards~\cite{ashforth2014dirty}. This category differs significantly from the other two; while physical and social dirty work are often seen as ``necessary but undesirable,'' moral dirty work is considered ``more evil than necessary.''~\cite{ashforth1999can,margolis2008navigating}. Many occupations in this category, such as gambling, sex work, and private investigation, are either socially unacceptable or even illegal in many regions. Therefore, the scope of this study excludes moral dirty work.

Research has shown that the stigma associated with dirty work not only leads to identity crises and significant stress for workers ~\cite{miscenko2016identity,ashforth2014dirty}, but also triggers negative emotional responses such as disgust and shame ~\cite{bolton2005women,simpson2012introducing}. Workers face discrimination from the public, as well as from colleagues and family members, which can lead to various physical and mental health issues. Moreover, this stigma results in organizational inefficiencies, high turnover rates, and increased social alienation and retaliatory behaviors~\cite{yan2017dirty}. It can also cause dirty workers to develop self-defeating social withdrawal mechanisms, further limiting their ability to engage in normal social activities ~\cite{abel2011different}.

Mainstream research on coping with the stigma of dirty work has often focused on the workers themselves, such as fostering a ``strong'' occupational and group culture to create a ``social buffer zone'' ~\cite{e2007normalizing} or developing positive occupational ideologies. However, since dirty work is primarily a social construct that depends on subjective perceptions ~\cite{emerson1976dirty}, a more fundamental approach may involve reshaping social understanding and empathy towards dirty workers, thereby adapting social views on these occupations.

Building on this foundation, we adopt the established definition of dirty work proposed by Ashforth et al.~\cite{ashforth1999can}, and select janitors, firefighters, police officers, and caregivers as representative occupations. Considering the relative scarcity of research on dirty work within the Chinese sociocultural context \cite{yan2017dirty}, we continue to employ this traditional theoretical framework in our study. We also acknowledge that prior research highlights the absence of unified standards for the classification and measurement of dirty work \cite{yan2017dirty}, and that stigma, due to its inherently subjective and socially constructed nature, is difficult to quantify directly.
Based on prior conceptualizations of stigma and dirty work \cite{ashforth1999can,goffman2009stigma,link2001conceptualizing}, we further conceptualize the stigma faced by dirty workers as a dynamic social process in which individuals or groups are negatively labeled, stereotyped, and marginalized through mechanisms such as labeling, stereotyping, and social exclusion. This process positions dirty workers in opposition to socially accepted ideals of cleanliness and respectability, symbolically associating them with moral, physical, or social impurity, and transferring these stains onto their identities. This conceptualization provides a theoretical foundation for understanding how social stigma toward dirty work is produced and maintained, thereby framing the significance of empathy-driven interventions for reducing such bias.

\subsection{Perspective-taking in Stigma Reduction}
Perspective-taking was defined as ``the process of imagining the world from another's vantage point or imagining oneself in another's shoes.~\cite{galinsky2005perspective}'' Gasiorek et al.~\cite{gasiorek2017perspectives} conceptualize perspective-taking as a situated process through which individuals form ideas about the content of another person's mental state. This process is supported by two key factors: (1) a social cognitive ability with a developmental trajectory, and (2) a general tendency for spontaneous engagement, which varies as an individual difference.

A substantial body of research has already demonstrated that perspective-taking offers a wide range of benefits. Engaging in the process of adopting another person's perspective has been shown to enhance liking, compassion, and willingness to provide assistance to the target individual~\cite{goldstein2014perceived,epley2006perspective,maner2002effects,vorauer2013potential}. Additionally, it fosters social interactions, facilitates smoother and more coordinated interpersonal communication, and aids individuals in achieving their objectives in strategic interactions such as negotiations~\cite{chartrand1999chameleon,galinsky2008pays,gilin2013use}. One particularly function of perspective-taking is its ability to reduce stigma and prejudice, not only toward the specific individual whose perspective is taken but also more broadly toward the individual’s social group~\cite{vescio2003perspective,galinsky2004effects,goldstein2014perceived}. Scaffidi et al. ~\cite{scaffidi2020effect} provides evidence that perspective-taking can effectively reduce group bias in language. Margaret et al.~\cite{maner2002effects}'s research revealed that perspective-taking not only improves overall perceptions of outgroups but also promotes positive behaviors at the individual level. Martingano et al.~\cite{martingano2023using} demonstrated that perspective-taking effectively reduces weight stigma and increases participants' empathy. Empathy has long been recognized as a powerful mechanism for reducing stigma toward stigmatized groups as a whole \cite{batson1997empathy}. Beyond its affective impact, empathy also serves as an effective strategy to overcome cognitive barriers, prompting individuals to critically reexamine and potentially revise their negative stereotypes about stigmatized populations \cite{clinton2021using}.

Previous studies have typically employed methods such as reading texts~\cite{freeman2015wonder}, watching videos~\cite{martingano2023using}, listening to radio broadcasts~\cite{batson1997perspective}, or engaging in virtual reality (VR) experiences~\cite{herrera2018building} to facilitate perspective-taking. However, no research to date has explored the use of LLMs for real-time, personalized interactions with users to create more dynamic and interactive experiences. This study aims to address this gap by investigating how LLMs can be leveraged to enhance perspective-taking through individualized, adaptive engagement.

\subsection{LLMs in interactive fiction}
Interactive fiction (IF) can be understood as ``books without pages'' or as ``commingled bits'' of information, as described in Being Digital~\cite{seegert2009doing}. The experience of IF offers two critical elements that are often absent in contemporary storytelling media such as books, theater, films, and television: autonomy, the ability to act and effect change, and interactivity, the ability to listen, think, and respond intelligently to the users ~\cite{stern1998interactive}. The strengths of IF lie in its ability to trigger participants’ imagination through linguistic symbols, creating a sense of immersion while empowering readers to actively engage with and influence the narrative. Its nonlinear structure and interactivity enliven virtual worlds, stimulate curiosity and deep reflection, and emphasize the core relationship between readers and texts, as conceptualized in literary theory under ``reader-response criticism''~\cite{stern1998interactive,tavinor2005videogames}.

IF has also found applications in other fields. In education, IF has been demonstrated to effectively empower learners, spark their interest, and foster creativity~\cite{kozdras2006interactive,flynn2019use}. In health, IF has shown potential in raising awareness about sexual health among college students and improving their capacity for behavioral change~\cite{londino2011decisions}. Consequently, IF holds promise as a perspective-taking tool to help the general public understand the experiences of workers engaged in dirty work. The rise of LLMs further enhances the potential of games by enabling the creation of more dynamic and engaging content. Previous studies have demonstrated that LLMs can generate dynamic and natural non-player character (NPC) dialogues in role-playing games (RPGs), significantly improving player immersion ~\cite{gallotta2024large,jordan2024retrofitted,huang2024generating}. 

As a game type that relies purely on textual presentation, IF has also shown promising prospects with the integration of LLMs. PsychoGAT introduced an innovative LLM-based psychometric paradigm by transforming traditional psychological questionnaires into interactive virtual novel games, enhancing both user engagement and the effectiveness of psychological assessments ~\cite{yang2024psychogat}. Similarly, ReZork leveraged LLMs to reimagine the classic IF Zork, creating a more seamless and immersive narrative experience ~\cite{tikhonov2024zork}. Notably, while previous studies have explored the use of LLMs to reduce stigma, they predominantly employed chatbot-based interventions\cite{lee2023exploring,cui2024exploring,reategui2025llm}. In contrast, our use of an IF framework represents a novel exploration of utilizing LLMs to reduce stigma, offering a fresh perspective on their application in promoting fairness.

\section{Method}
While previous research has explored the potential of perspective-taking and LLM in reducing stigma, there remains a gap in their application within immersive, interactive formats like IF. Building on these insights, our study leverages an LLM-based IF framework to simulate the daily experiences of dirty workers, encouraging perspective-taking through real-time, adaptive interactions. The following section details the design of this approach and its implementation to address our research questions.

\subsection{LLM-based IF for perspective-taking}
To explore RQs, we aimed to design the LLM-based perspective-taking IF. This IF would immerse users in the daily experiences of dirty workers, encouraging perspective-taking and fostering a deeper understanding of the challenges they face. To enable a LLM to construct scenarios for this IF and assist participants in better understanding dirty workers through perspective-taking, we initially drew on prior research~\cite{ho2022perspective,herrera2018building,clinton2021using} and design principles for prompt engineering~\cite{chen2023unleashing,marvin2023prompt}. The specific prompts and their design principles can refer \autoref{appendix:prompt}. We developed IF representing four different occupations on the Poe platform\footnote{https://poe.com}. GPT-4o was selected as the base model. The entire user experience was structured in three main phases: an introductory guide, scenario description and user interaction, and a summary reminder.

The introductory guide serves to introduce the occupation being simulated by the LLM and provide instructions on how users can begin their experience. The scenario description and user interaction constitute the core of the perspective-taking experience. In this phase, users step into the shoes of a dirty worker for a day, experiencing a series of potential work scenarios that they might face. Following each scenario, users are presented with several example decision options, ensuring that, as non-professionals, they have guidance on how to respond. However, users also have the option to deviate from these suggestions and input their own responses. Once a choice is made, the LLM provides real-time feedback, showing the possible consequences of their decision and advancing the narrative based on the user’s choices. Each scenario does not present a singular outcome but leads to new challenges, depending on the user’s decisions, simulating a more dynamic and realistic flow of events until a resolution is reached. Based on the Ease of Self-Simulation heuristic (ESS) , the design incorporates detailed background descriptions and linguistic cues to help participants more readily imagine themselves in the roles of dirty workers ~\cite{chambers2012role}. This approach serves two purposes: first, it avoids overly simplistic scenarios that are easily resolved, making the situations more reflective of real-world complexities; second, it enhances elements of resonance with daily life, fostering a stronger sense of participant engagement ~\cite{chambers2012role}.

To provide a more nuanced perspective, 1–2 scenarios depicting challenges faced by dirty workers in their personal lives are introduced alongside the work-related contexts. For example, a police officer may be too busy to remember a family member’s birthday, or a firefighter might face criticism from loved ones after sustaining injuries. These scenarios aim to give participants a broader understanding of the personal challenges that dirty workers might encounter outside of their professional duties. Additionally, the design incorporates potential biases or stigmas faced by dirty workers in their work and personal lives, as generated by the LLM.  According to the Social Cognitive Theory(SCT), observational learning involves acquiring attitudes, values, and cognitive and behavioral patterns by observing examples set by others~\cite{bandura2008observational,bandura2001social}. Participants, by simulating these scenarios, can understand the difficulties faced by dirty workers in their personal lives and the complexities of the social responses behind them. These scenarios not only demonstrate the impact of job-related stress on personal life but also address the social evaluations of others regarding the behaviors or choices of these workers. The presence of such social evaluations and biases can prompt participants to reflect on the social prejudices against dirty work professions through observing others' actions and their consequences. Based on the concept of Reciprocal Determinism in SCT, individuals are influenced by the external environment, while their behaviors also affect the environment, and their cognitive factors influence both their behavior and their responses to the environment~\cite{schiavo2019dynamical,bandura2001social}. By allowing participants to experience the impact of their actions on simulated scenarios through an IF, we aim to influence participants and their cognitive processes, thereby reducing stigma and bias.

At the end of the experience, the LLM provides a summary that reflects on the user’s choices throughout the day and the nature of the work. The summary emphasizes that the scenarios presented are just a small part of the daily reality for dirty workers, and reminds users that, in real-world contexts, workers often face more complex and unexpected situations. The model also highlights the skills and competencies these workers need to navigate such challenges. Additionally, the summary offers prompts for further reflection, such as, ``If you were an employer, how would you better support a caregiver and acknowledge their struggles?'' This structure is designed to foster deeper empathy and a broader understanding of the multifaceted experiences of dirty workers.

In addition, we invited workers from these occupations to evaluate the IF and further refined the prompts based on their feedback. For more details, please refer to the \autoref{appendix:1}  and \autoref{tab:occupations}.

\subsection{Procedure and Survey Design}

We recruited 100 participants (Average age = 25.5; 57 male, 43 female) through specific institutional online communities, such as university forums and internal research group platforms, and targeted email outreach to 127 potential candidates identified through relevant mailing lists and online advertisements. All participants were native Chinese speakers residing in China. Eligibility required English proficiency to enable post-interview verification of translated responses through direct participant consultation, alongside age $\geq$ 18 years and no prior experience in the four target occupations. Participants received standardized compensation (e.g., equivalent to an hourly rate for their time, with increased compensation for those who engaged in follow-up interviews) and were randomly assigned to four groups (n = 25 per group) exposed to distinct occupational simulations (janitors, firefighters, police officers, and caregivers). Pre-experimental procedures included detailed ethical briefings on opt-out mechanisms and withdrawal rights and completion of baseline surveys. The survey included the Interpersonal Reactivity Index (IRI)~\cite{davis1983measuring} to assess individual empathy traits, and a Beliefs about Empathy ~\cite{schumann2014addressing} measure to determine whether participants viewed empathy as a controllable trait and whether they subscribed to a fixed'' or malleable'' theory of empathy. These assessments ensured that empathy traits were balanced across the four groups, with no significant differences between them.

Additionally, as previous research has shown that increasing awareness and knowledge of stigmatized groups can reduce prejudice~\cite{hinshaw2000stigma}, we incorporated a custom scale for participants to complete (\autoref{appendix:scale}). This scale was specifically designed to measure participants’ prior knowledge and perceptions of the particular dirty work profession they were about to experience. After the pre-test, participants could use tools such as AI face-swapping software to merge their own photo with an image representing the assigned dirty work, enhancing their sense of immersion. To protect privacy, we did not collect these images. Next, participants engaged in the IF related to their assigned dirty work. Upon completing the session, participants exported the conversation link and submitted it to us.

In the post-test phase, participants completed several measures. The Empathy and Personal Distress scales~\cite{batson1997perspective} were used to assess participants’ empathetic reactions as well as their levels of tension, discomfort, and distress during the task. Measuring empathy is essential, as prior research has demonstrated that empathy serves as a powerful mechanism for reducing stigma—both by promoting more positive attitudes toward stigmatized groups~\cite{batson1997empathy} and by facilitating the revision of negative stereotypes through the mitigation of cognitive barriers~\cite{clinton2021using}.

While the Social Distance Scale~\cite{bogardus1933social} has commonly been used in previous research to quantify the degree of social contact, we argue that it is more appropriate for examining more severe forms of social exclusion or stigma. Informed by prior work~\cite{herrera2018building}, we instead employed the Inclusion of the Other in the Self (IOS) scale~\cite{aron1992inclusion} to assess participants’ perceived closeness to dirty workers following the experience. The IOS scale, grounded in Self-Other Overlap theory~\cite{waugh2006nice}, offers a more nuanced and sensitive means of capturing shifts in social identity and relational proximity resulting from empathy-driven interventions.

Finally, the custom-designed scale used in the pre-test was re-administered to evaluate whether participants’ knowledge and perceptions of dirty work occupations had changed. To further enrich our understanding, our qualitative analysis also explored participants' perceptions of these occupations before and after the intervention, allowing us to triangulate the data and gain a more holistic view of how empathy influences stigma and stereotypes.

\section{Results}
We present an overview of quantitative findings for each research question and contextualize our results with insights from the semi-structured interviews. To facilitate understanding, we include a complete example of a user's interaction flow during the Caregiver IF experience, referenced in \autoref{dialogue}.

\subsection{RQ1: Effectiveness of LLM-Based IF in Reducing Dirty Work Stigma}
First, using the results from the IRI and Beliefs About Empathy scales, we conducted an ANOVA to assess significance. The analysis revealed no significant differences in empathy levels among participants across the four experimental groups 
 (can see in \autoref{tables}: \autoref{tab:IRI}). This indicates that the experimental results were not influenced by differences in participants' empathy traits, eliminating this potential source of bias. Post-experiment results revealed that participants scored highly on both empathy and distress measures (can see in \autoref{tables}: \autoref{tab:empathy-distress-ios}), indicating that they developed empathy for the corresponding dirty workers and experienced their distress. The high scores on the IOS scale further demonstrate that participants formed closer connections with the dirty workers. Importantly, no significant differences were found across the four experimental groups on these three scales (can see in \autoref{tables}: \autoref{tab:empathy-distress-ios}), suggesting that participants’ responses were consistent regardless of the specific dirty work profession. To further validate the consistency of responses across different occupational groups, we conducted multiple comparisons using Bonferroni correction (\autoref{tables}: \autoref{tab:bonferroni_updated}). The results indicated that no statistically significant differences were found between any pairs of occupational groups across all five measures: Beliefs About Empathy, IOS, IRI, Empathy, and Distress (all adjusted p-values > 0.05). This finding provides additional confirmation that participants’ emotional reactions and relational closeness toward dirty workers were consistent regardless of the specific occupation presented. It also reinforces the robustness of the experimental effects by ruling out potential confounding influences arising from differences between occupational groups. For our custom-designed scale, we used a Paired t-test to compare the significance between the pre-test and post-test. Results showed significant differences in all categories except for the occupational value perception of police officers and firefighters, as well as the occupational stress and risk perception of firefighters(can see \autoref{tables}: \autoref{tab:caregivers}, \autoref{tab:police_officers}, \autoref{tab:firefighters}, \autoref{tab:janitors}).  These results suggest that participants' understanding and awareness of most aspects of dirty work professions improved significantly after the experiment. The lack of significant differences in the occupational value perception of police officers and firefighters aligns with dirty work theory~\cite{yan2017dirty}. These professions, being highly socially honored and positively portrayed, tend to carry greater social recognition and positive reputations. Similarly, due to extensive positive media coverage and widespread reports about firefighters, the public may already have a relatively high level of awareness regarding the pressures and risks associated with their work. However, even in these cases, participants’ understanding in these areas still improved to some extent following the experiment.

In summary, these findings suggest that LLM-based perspective-taking IF can help reduce stigma toward dirty workers. This approach not only enhances participants' understanding of these professions but also fosters empathy and deeper connections, thereby addressing social biases and promoting greater awareness.

\subsection{RQ2: Mechanisms of LLM-Based Perspective-Taking}

\begin{table}[ht]
\centering
\caption{Demographic and Professional Information of Interview Participants. The column labeled \textit{Experience} refers to participants' experience with "dirty work" professions.}
\label{tab:dirty_work}
\resizebox{\columnwidth}{!}{%
\begin{tabular}{@{}l c c p{4cm} p{2cm}@{}}
\toprule
\textbf{ID} & \textbf{Gender} & \textbf{Age} & \textbf{Profession} & \textbf{Experience} \\
\midrule
P1  & Female & 22 & Sales                 & Firefighters       \\
P2  & Male   & 23 & Student               & Caregivers         \\
P3  & Male   & 24 & Programmer            & Police Officers    \\
P4  & Male   & 22 & Student               & Janitors           \\
P5  & Male   & 24 & Accountant            & Firefighters       \\
P6  & Male   & 25 & Student               & Firefighters       \\
P7  & Female & 23 & Freelancer            & Janitors           \\
P8  & Female & 24 & Student               & Police Officers    \\
P9  & Female & 23 & Public Opinion Analyst & Caregivers         \\
P10 & Female & 23 & Student               & Caregivers         \\
P11 & Male   & 25 & Student               & Janitors           \\
P12 & Female & 23 & Designer              & Caregivers         \\
P13 & Female & 28 & PM New Media Operations & Police Officers    \\
P14 & Male   & 32 & Research Assistant    & Firefighters       \\
P15 & Male   & 30 & Front-end Developer   & Firefighters       \\
\bottomrule
\end{tabular}%
}
\end{table}

To further investigate why LLM-based perspective-taking IF can effectively reduce stigma toward dirty work, we conducted semi-structured interviews with participants (n=15), all of whom agreed to participate. The interviews were conducted via Zoom, following a semi-structured format that focused on exploring participants' emotional responses, their attitudes toward LLM-based perspective-taking IF, and their insights into how this approach influenced their perceptions of dirty work. Example interview questions can be found in \autoref{tab:example}, and detailed information about the interview participants can be found in \autoref{tab:dirty_work}. Each Zoom interview was recorded and then transcribed verbatim to ensure the accuracy and richness of the participants’ subjective experiences. The first round of open coding \cite{corbin2014basics}, based on interpretative phenomenological analysis \cite{smith2007hermeneutics}, identified meaningful units related to LLM-based perspective-taking and its effect on attitudes toward dirty work. During this process, these units were extracted and assigned to specific codes. Subsequently, a second round of coding was performed, categorizing the initial codes into broader themes related to the mechanisms of LLM-based perspective-taking. Finally, all data, codes, and themes were translated into English for reporting purposes.

\subsubsection{Perceptions and Experiences of Professions in Context} 
Participants reported that using the LLM-based perspective-taking tool enhanced their comprehensive understanding of these ``dirty work'' professions.

\textit{\textbf{Empathizing with Job Responsibilities.} }
Participants expressed that the IF provided them with a deeper understanding of the daily tasks and responsibilities involved in ``dirty work'' professions. Due to their lack of direct exposure to these occupations or their practitioners in their daily lives and work, many participants had limited prior knowledge of such roles. This comprehensive perspective-taking allowed them to gain genuine insights into the everyday duties of these professions. \textit{``Previously, my understanding of police work was purely based on imagination—I had no idea what their actual responsibilities were. It was limited to what I learned from primary school textbooks, like ‘police catch bad people,’ without any concrete understanding. However, this contextual experience gave me a step-by-step sense of their entire day’s work. It felt much more profound compared to my prior perceptions.''}(P13)
The experience also enabled participants to move beyond their preconceived notions or stereotypes, fostering a more nuanced and comprehensive understanding of these professions. \textit{``I used to think firefighters just put out fires, but after this experience, I realized their work goes far beyond that. It truly helped me understand the breadth and variety of their responsibilities.''}(P15)
At the same time, the experience helped participants recognize that some jobs may require a variety of skills:  
\textit{``In my impression, janitors just performed simple tasks like sweeping or cleaning. However, after the experience, I could clearly see that their work is far from simple. They also need other skills, such as problem-solving on the spot and effectively communicating with pedestrians, coworkers, or others they interact with. '' (P4)}

\textit{\textbf{Interpreting Work Environments.} }
In addition to a lack of understanding regarding the tasks and responsibilities of these ``dirty work'' professions, participants also expressed unfamiliarity with the work environments. They were often unaware of the primary settings and contexts in which these workers operate. Through comprehensive simulation experiences, participants were able to gain a better understanding of these environments.\textit{``As a caregiver, I had to take care of children or elderly people in someone else’s home, and possibly help with cooking and housework. Being in a stranger's house and taking orders felt very oppressive.'' (P2)}\textit{``After the experience, I realized that firefighters have to deal with many hazardous situations. Compared to other professions, their working environment is significantly more dangerous.'' (P1)}

\subsubsection{Exploring Occupational Pressures and Associated Risks} 
Participants stated that after the experience, they developed a deeper understanding of the multifaceted pressures and potential risks associated with these jobs.

\textit{\textbf{Navigating Pressures from Internal Job Demands.} }
These dirty workers often face significant occupational stress\cite{ashforth1999can,ashforth2014dirty}. Due to limited or biased understanding of these professions, the public may remain unaware of the pressures they entail. Increasing awareness of these challenges could help reduce social misunderstandings about these occupations. For instance, P14 observed that the work of firefighters is characterized by high levels of unpredictability, which often disrupts even basic routines such as meal times. \textit{``When I’m about to eat, an emergency might suddenly arise. Should I respond to it? Of course, I must go immediately because even a slight delay could lead to serious consequences.'' (P14)} Another participant shared a memorable experience: \textit{``One experience that left a deep impression on me was receiving a call after finishing my shift late at night. Although I had already been busy all day, I had no choice but to handle the situation.'' (P3)} Participants also reported experiencing the complexity inherent in their professions, which contributes to significant stress. \textit{`` When multiple tasks come at you simultaneously, it’s difficult to determine the optimal order of action. Just thinking about how to respond has already been quite overwhelming for me. Imagining how challenging it would be to perform these tasks as a caregiver in real-life scenarios makes me feel it could be even harder.'' (P9)}  \textit{``It feels like caregivers have to deal with a lot of complicated and frustrating matters. Often, many tasks pile up at the same time. When conflicts arise, it’s impossible to satisfy everyone.'' (P12)}

\textit{\textbf{Examining Pressures from Family Role Conflicts.}} 
In addition to occupational stress, these dirty workers also face pressures stemming from family dynamics. Beyond the stress of the work itself, participants noted that familial pressure and lack of understanding can exacerbate their emotional burden. \textit{``After experiencing this, I felt that working hard all day to care for other families and then coming home at night, not only unable to rest but also facing a lack of understanding from my own family.'' (P2)} Moreover, the demanding nature of these jobs can lead to workers forgetting or missing significant family occasions, such as birthdays or prearranged activities. This not only strains relationships with family members but also places additional psychological pressure on the workers. \textit{``Police officers are ordinary people too; they have families and lives. However, their work often separates them from their families. For instance, when faced with the choice of celebrating my child’s birthday or staying on duty, while I chose to celebrate the birthday during the simulation, in reality, I might have no choice but to remain on the front line.'' (P3)} Another participant observed, \textit{``Due to the nature, urgency, and specificity of this work, conflicts between work and family are magnified significantly. These conflicts create additional challenges. I feel that the most critical and difficult aspect is balancing work and family life.'' (P5)}

\textit{\textbf{Understanding Pressures from Public Judgments.} }
Since these occupations are labeled as ``dirty work,'' social judgments and stigmas are often projected onto the workers themselves, making it unavoidable for them to face negative evaluations and disapproval in the course of their duties~\cite{ashforth1999can}. \textit{``I think this profession suffers from the most severe stigmatization. As mentioned in the simulation, many parents and teachers often say things like, ‘If you don’t study hard, you’ll end up sweeping the streets.’ This kind of remark completely devalues the worth of the profession.'' (P7)} Another participant echoed this sentiment: \textit{``From a young age, many parents might use sweeping the streets as a way to scare children into studying harder. This creates stereotypes that are deeply ingrained. After experiencing this perspective shift, I can genuinely feel how harmful such remarks can be.'' (P11)}In addition to these biases, social evaluations can directly impact the execution of daily work. \textit{``When I was mediating a conflict, someone in the crowd was making comments and watching for amusement. If it were me, I would definitely be distracted by this.'' (P3)} \textit{``When handling an emergency, if the situation is not resolved properly, the worker might face criticism and misunderstandings from others. This is indeed challenging and really not easy.'' (P15)} Furthermore, social judgments can deepen the emotional divide between the worker’s professional life and the sense of honor associated with their occupation. As described by P5, \textit{``I noticed how the options in the simulation could lead to property losses, compounded by negative evaluations from others. It highlighted the gap between the realities of this profession and the sense of pride it should carry.'' (P5)}

\textit{\textbf{Exploring Occupational Risks via Life-Threatening and Social Impacts} }
One characteristic of dirty work, particularly for firefighters and police officers, is the inherent danger involved~\cite{ashforth1999can}. \textit{``For the first time, I felt what it might be like for someone to rush into a blazing fire to save others. It really felt incredibly dangerous. Even though they are fully equipped, it seems useless in the face of such a massive fire.'' (P5)} Additionally, these roles often entail social risks, distinguishing them from ordinary occupations. Participants deeply experienced this aspect during the simulation: \textit{``This time, I truly realized that being a caregiver is not only difficult but also risky. You have to handle so many tasks, and failing to do any of them properly could lead to criticism from employers.'' (P9)}Moreover, the unique nature of dirty work means that workers of different genders may face varying levels of danger. \textit{``I could feel the danger in this job. I remember being in a stairwell where I encountered a suspicious man. If I were a female police officer, I would have been extremely afraid to approach him because of the disparity in physical strength. But as a male, I felt more confident walking up to him and asking questions.'' (P13)} The decisions participants make in response can also lead to vastly different outcomes, posing varying degrees of risk. This aspect highlighted the difficulty and significance of decision-making in such professions. \textit{``It’s about striking the right balance. If the person turns out to be innocent, it could lead to public backlash. But if he is indeed a dangerous individual, prematurely alerting him might escalate the situation and lead to even worse consequences.'' (P13)}

\subsubsection{Enhancing Emotional Resonance and Self-Reflection} 
Participants noted that the experience heightened their empathy for ``dirty work'' practitioners and prompted profound self-reflection on their own perceptions and attitudes.

\textit{\textbf{Deepening Emotional Connection through Personal Relatability.} }
Certain scenarios in the simulation resonated deeply with participants, as they were reminded of their own personal experiences, fostering a stronger sense of empathy and connection. \textit{``In the simulation, there was a scene where my wife was upset, and I needed to comfort her. This really resonated with me because it reflects my own life experience—how I often try to soothe my partner’s emotions. There was also the part where my meal was interrupted, which I found very relatable. It felt like my life was being disrupted—like when I’m eating and suddenly there’s a problem that urgently needs my attention. I have to put down my chopsticks and address it immediately. That feeling is very unpleasant. Although it’s hard to describe the exact emotions, I found it to be extremely distressing.'' (P13)}In addition, some experiences in the simulation mirrored situations that are universally familiar, further enhancing the participants’ sense of connection. \textit{``For example, in the experiment, the employer’s plans changed, and they decided to take the child out for a meal. This reminded me of my boss, who might assign me a task and then suddenly change it, asking me to do something entirely different instead. It’s a similar feeling.'' (P10)}

\textit{\textbf{Developing Understanding through Role Immersion.} }
Through this comprehensive perspective-taking experience, participants gained a deeper understanding of these professions and developed greater empathy for the workers in these roles. \textit{``After gaining a profound understanding of the basic duties, challenges, and social responsibilities faced by firefighters, I am better able to empathize and put myself in their shoes to understand these professions.'' (P6)} Another participant shared, \textit{``After experiencing it, I felt a lot and became more familiar with the nature and content of their work. It significantly enhanced my empathy and helped me reduce many of the biases I previously held against them.'' (P4)}The experience also increased participants’ understanding of the behaviors of these workers in everyday life. \textit{``Janitors might face significant stress, so sometimes their attitude while cleaning or performing their tasks may not be entirely positive. By enhancing empathy or engaging in perspective-taking, one might better understand the reasons behind their less favorable attitudes.'' (P11)} Moreover, the simulation brought participants closer to dirty workers, bridging a gap that had previously seemed insurmountable. \textit{``Before, without specific experiences or scenarios, I felt that police officers were very distant from me. But after truly experiencing their role, I felt much closer to them, which fostered a sense of empathy.'' (P3)}

\textit{\textbf{Reinforcing Learning through Guided Reflection.}} 
At the conclusion of the experience, a summary session was provided, which received positive feedback from participants. \textit{``There was a summary at the end, and I thought it was excellent. Without this final summary, I might not have been able to grasp all the details or might have missed some things. I might forget many of the small details, but those seemingly insignificant small tasks actually add up to a considerable workload.'' (P14)} The LLM also provided participants with a summary of the day’s activities and posed several reflective questions, which further encouraged introspection. \textit{``The final summary and reflection were quite good. It gave me the opportunity to spend time reflecting on the day and to review it from the perspectives suggested by those questions.'' (P5)}

\subsubsection{Fostering Experiential Immersion via Proactive Participation} 
Participants felt that this immersive approach allowed them to step into the workers' shoes, deeply experiencing their daily routines and fostering a strong sense of involvement.The use of IF with a first-person perspective allows participants to engage directly with scenarios and make their own decisions about the situations they encounter, providing them with a stronger sense of immersion and experiential depth. \textit{``Previously, I viewed stories from a third-person perspective, but experiencing them in the first person provided a much deeper understanding. Different perspectives result in very different perceptions and feelings about the same event.'' (P3)} Additionally, the integration of LLMs offers participants greater flexibility and choice in their interactions. \textit{``LLMs are much more flexible. Compared to pre-set questions and fixed scenarios, participants can chat freely and express their own thoughts to the LLM. The LLM then adjusts its responses accordingly. I think this makes the experience even more immersive.'' (P4)} Furthermore, the unique format of IF enhances participants' sense of involvement. \textit{``It feels like a conversation, like a text-based game. The system adjusts in real-time based on user input. This approach is really interesting—it allows you to experience a life you've never lived before.'' (P1)}

\subsubsection{Promoting Respect for On-Site Professionals} 
Participants believed that the immersive experience and deeper understanding enhanced their respect for workers in these fields.The experience further deepened participants’ understanding of the challenges faced by these workers, fostering greater respect for these professions.\textit{``I have always had a basic level of respect and admiration for the firefighting profession. However, through the simulation, I gained a deeper understanding of the nature of their work, which inherently requires them to ‘sacrifice their own families for the greater good,’ as well as the risks involved in their duties.'' (P6)} Beyond gaining respect for the professions themselves, the experience also prompted participants to reflect on the importance of occupational equality. \textit{``We should indeed respect them. They work very hard and endure significant fatigue. In fact, there should be no hierarchical distinctions among professions. I believe we should treat all occupations with equal respect and regard.'' (P4)}

\subsubsection{Exploring Pathways to Professional Fulfillment} 
Some aspects of the experience allowed participants to gain an appreciation for the sense of accomplishment and satisfaction that comes with dirty workers, helping them better understand the value of these professions.\textit{``My focus was on the scenario where a five-year-old child came home from school, and after I provided emotional guidance, the child seemed to release their emotions. This included processing things they didn’t understand and alleviating their negative feelings. I felt a sense of accomplishment from this.'' (P9)} Another participant shared, \textit{``Rescuing a small cat or receiving the praise of the public—something like that. It gave me a sense of small but genuine accomplishment.'' (P15)}

\subsection{RQ3: Challenges and Limitations of LLM-Based IF for Stigma Reduction}

\subsubsection{Limited Narrative Depth and Character Development} 
Although we utilized one of the most advanced LLMs available, GPT-4o, as the foundation and employed highly detailed prompts for functionality settings, some participants felt that the perspective-taking IF scenarios generated by the LLM still lacked detail in certain aspects. \textit{``I think the generated content is not comprehensive enough. It needs to simulate more details. However, this is primarily because the LLM itself needs to be more advanced.'' (P3)} Additionally, participants expressed a desire for richer narratives and more engaging storylines. \textit{``Perhaps my expectations are high since I’m comparing it to games like Baldur’s Gate. I think the storylines could be made richer, and elements like supporting characters and side quests could be added, such as developing plots involving teammates.'' (P5)}

\subsubsection{Unintended Amplification of Occupational Stereotypes} 
Participants noted that certain storylines in the simulation might inadvertently reinforce stereotypes about these professions, a phenomenon consistent with findings in previous stigma-related studies, such as the concept of suppression—where efforts to suppress an emotion or stereotype may paradoxically amplify the very emotion or stereotype intended to be minimized~\cite{hinshaw2000stigma,corrigan1999lessons}. Similarly, this also mirrors the 'paradox of empathy,' wherein empathy may paradoxically elicit adverse emotional reactions\cite{meugnot2024navigating}. As one participant explained, \textit{``The simulation does highlight the dangers of firefighting, which might intensify the public’s stereotypes. While it may indeed increase empathy, it could also make people more reluctant to become firefighters themselves or to encourage their family members to take up such roles. These two outcomes are not mutually exclusive.'' (P1)}

\subsubsection{Requirements for User Engagement} 
While the IF format itself was well-received by participants, they felt that in an era of advanced technology, text-based interactions alone might not be sufficient to fully engage users. They suggested incorporating additional modalities and more engaging content and interaction methods to enhance the experience. \textit{``Text alone feels somewhat limiting. It primarily relies on triggering your imagination. However, if there were visual elements to guide the experience, I think the immersion would be much better.'' (P14)} Another participant echoed this sentiment, \textit{``Currently, ChatGPT is still text-based, which I feel might not provide enough immersion. Future iterations could, for instance, generate images or even integrate with MR or VR devices to create a more immersive experience.'' (P15)}

\section{Discussion}
Our study explores the use of LLM technology to construct an innovative perspective-taking IF to reduce stigma associated with dirty work, a unique category of occupations, particularly focusing on physical dirty work and social dirty work. Our experiment found that LLM-based IF helps participants reduce stigma toward dirty workers (RQ1). Additionally, we investigated the mechanisms through which this approach operation (RQ2), as well as its challenges and limitations (RQ3). Our research aims to leverage emerging technologies to promote social and occupational equity, taking a step toward a more harmonious culture.

\subsection{Generalizability of LLM-Based Interactive Fiction for Reducing Stigma}
Our findings demonstrate that LLM-based perspective-taking IF is an effective approach to reducing occupational stigma associated with dirty work \textbf{(RQ1)}. Quantitative comparisons of participants' pre-test and post-test measures revealed a significant increase in understanding of dirty work. Meanwhile, the participants demonstrated high levels of empathy and distress, as well as a closer perceived proximity to dirty work professions. These results align with and extend previous research, highlighting the benefits of perspective-taking in reducing bias and stigma ~\cite{vescio2003perspective,galinsky2004effects,scaffidi2020effect}. Notably, while earlier interventions relied on relatively static content ~\cite{batson1997perspective,martingano2023using}, the LLM-based approach enables dynamic, customizable scenarios and responses, allowing participants to engage more deeply with the challenges faced by workers in their daily lives. This adaptability permits more personalized and context-rich interactions, effectively fostering empathy.

Beyond its efficacy, this approach is also accessible and scalable. Participants accessed scenarios through an online chatbot interface, enabling broader dissemination to the public. In addition, participants noted the privacy advantages of this approach. By simulating fictionalized scenarios, the method protects the privacy of real workers, such as police officers who may handle sensitive or dangerous cases. This balance of accessibility, immersion, and privacy protection fulfills calls in the literature for more flexible and immersive technologies to reduce social biases ~\cite{herrera2018building}. Thus, our study introduces a novel method that enriches digital interventions aimed at stigma reduction and has potential applications for other socially devalued or marginalized roles.

While results align with established perspective-taking mechanisms, three directions will strengthen generalizability: 1) Longitudinal follow-ups assessing empathy/stigma persistence across time intervals, 2) Durability strategies through repeated scenario exposure and reinforcement mechanisms, and 3) Expansion to diverse populations spanning age groups, cultural contexts, and community types. These extensions will test the approach's adaptability to broader social contexts while preserving its core immersive advantages.

\subsection{Mechanisms Underlying the Reduction of Occupational Stigma}
Our qualitative findings \textbf{(RQ2)} reveal why LLM-based IF effectively reduces the stigma of dirty work. First, by gaining a comprehensive understanding of job responsibilities, work environments, and the complex tasks involved in dirty work, participants experience a cognitive shift that aligns with the principles of social cognitive theory~\cite{luszczynska2015social}. This deeper awareness not only reshapes their perceptions but also influences behaviors, ultimately counteracting stereotypes rooted in superficial or outdated impressions. Second, through meaningful exposure and interaction, participants gained a nuanced appreciation of the immense stress and risks these workers endure in both their professional and personal lives—including physical hazards, social stigma, and family conflicts. This recognition aligns with the contact hypothesis, highlighting how direct engagement can foster empathy and challenge preconceived notions~\cite{paluck2019contact,amir1969contact}. By exposing the demands on dirty workers, the IF experience enabled participants to empathize with the emotional and psychological burdens these workers bear. Third, the immersive and interactive nature of LLM-based IF enhanced users' emotional resonance. Participants freely navigated branching scenarios and experienced personalized consequences, fostering self-reflection and heightened empathy. Finally, the IF format allowed participants to appreciate the sense of accomplishment and fulfillment inherent in these professions, demonstrating that despite their challenges, dirty work is indispensable and provides personal rewards such as pride and achievement. These findings underscore the power of combining perspective-taking with adaptive, text-based simulations to evoke empathy, dispel misconceptions, and promote respect. They also highlight how cultivating ``walking in others' shoes'' experiences can lead to lasting attitude changes.

\subsection{Empowering Through LLMs: Mitigating Challenges in Stigma Reduction}
Based on our findings in \textbf{RQ3}, we identified several challenges faced in implementing LLM-based perspective-taking IF, which highlight both the limitations of current technologies and their transformative potential in addressing stigma. These challenges—ranging from narrative depth and stereotype amplification to technical engagement constraints—underscore the complexity of using artificial intelligence to foster empathy and reduce bias. 
Firstly, participants encountered narrative depth limitations due to current LLM constraints, particularly in generating nuanced and contextually rich scenarios. This was particularly evident among participants less familiar with the occupational contexts depicted, who found the narratives overly simplistic. For example, while P13 found the scenarios relatable, P3 thought the occupational portrayals lacked detail. Secondly, participants may inadvertently reinforce existing stereotypes through oversimplified occupational portrayals, as discussed in Section 4.3. Participants often focused on surface-level challenges, such as the dangers of firefighting, while overlooking broader aspects like professional skills or societal contributions. This paradox highlights the delicate balance between addressing hardships and avoiding stereotype amplification~\cite{hinshaw2000stigma,corrigan1999lessons}. Thirdly, we observed that participants use text-based interactions to understand occupational perspectives. While these non-immersive approaches can somewhat bridge empathy gaps—particularly benefiting users without access to advanced technologies—they may limit emotional engagement and long-term impact. 

In fact, technological advancement holds powerful potential to mitigate these challenges. LLMs' architectures and multimodal integration can bridge experiential gaps and address narrative limitations by providing richer, more nuanced content\cite{xing2024survey}. Without such technological evolution, users often had to rely solely on text-based interactions without supplementary sensory engagement. Multimodal technology facilitates deeper immersion, enhances narrative authenticity, and fosters stronger emotional connections between users and simulated experiences \cite{sakr2016mobile,xu2022understanding,flavell2013perspectives}. Besides, such LLM models capabilities allows for more sophisticated simulations. Users can engage with more complex scenarios and potential outcomes, enabling informed perspective-taking and deeper understanding \cite{zhang2024llm,li2024more}. For instance, users could experience multiple viewpoints of the same occupation, leveraging diverse narratives to share experiences and challenge stereotypes. 

To address challenges in LLM-based perspective-taking IF approach, we propose several future directions. First, personalized narrative generation based on user characteristics and cultural contexts can be beneficial. For instance, users might benefit from scenarios that bridge their experiences with unfamiliar occupational contexts, enhancing narrative relatability and emotional engagement. Second, integrating multimodal elements into LLM-based simulations can provide substantial benefits. Researchers could develop hybrid approaches combining text, visual, and interactive elements, enabling more immersive perspective-taking experiences while maintaining accessibility and scalability \cite{feldman2023trapping}. Additionally, advanced LLM architectures can serve as more sophisticated narrative generators, allowing for dynamic scenario adaptation and deeper character development. These models offer promising potential for creating more nuanced and culturally sensitive simulations \cite{wan2023kelly,kwok2024evaluating}. These advancements help mitigate current limitations in LLM-based perspective-taking, leading to more effective stigma reduction outcomes.

Beyond LLMs themselves, as P1 emphasized, the amplification of stereotypes stems not exclusively from LLM but reflects broader challenges in raising awareness about the risks and hardships linked to certain professions. While participants demonstrated increased empathy, this awareness paradoxically heightened their reluctance to pursue these professions. This phenomenon reflects the concepts of suppression and the paradox of empathy\cite{hinshaw2000stigma,meugnot2024navigating}. Future research will focus on strategically balancing empathy generation with preventing the reinforcement of negative perceptions, particularly through narrative design that promotes understanding and stigma reduction without unintentionally increasing professional reluctance or perpetuating stereotypes.

\subsection{Limitations}
Despite its promising potential, this study has several limitations. First, although individual differences in empathy were accounted for using measures like the IRI, generalizability may be constrained by cultural factors. As the sample was primarily drawn from China, where certain occupations (e.g., janitorial work) carry heightened stigma, findings may not extend to other sociocultural contexts. 
Future research could explore ethical approaches to reducing stigma for morally ambiguous professions. 
Second, the study focused on physical and social dirty work, excluding moral dirty work, which involves ethical complexities \cite{ashforth1999can}. Exploring stigma reduction for morally ambiguous professions presents an important avenue for future research. 
Third, reliance on self-reported measures may introduce bias, and the study did not assess the long-term effects of the intervention. Future work could adopt behavioral indicators, longitudinal designs, or experimental comparisons (e.g., text-based vs. immersive VR interventions) to evaluate durability and efficacy more comprehensively. 
Fourth, our study did not directly compare the effectiveness of LLM-based perspective-taking with other established stigma-reduction methods. While this research contributes to exploring a novel approach offering rich interactive experiences, future work should include comparative studies to determine the relative efficacy of LLM-based interventions versus traditional perspective-taking techniques and other stigma-reduction strategies. 
Fifth, despite our detailed prompt engineering efforts, some participants noted limitations in the richness and depth of LLM-generated narratives. This reflects current technological constraints in generating truly immersive occupational experiences. Future research may explore enhancements like Retrieval-Augmented Generation (RAG) to incorporate more authentic occupational details and specialized knowledge, potentially creating more convincing and nuanced simulations of dirty work experiences.
Lastly, the potential biases and inaccuracies of LLM-generated content remain a concern. Although rigorous prompt design minimized risks and no problematic content was reported, the uncertainty of current LLM technologies persists. Future research should prioritize improving LLM accuracy and developing monitoring protocols to ensure ethical and effective applications \cite{feldman2023trapping,wan2023kelly}.

\section{Conclusion}
This study demonstrates that LLM-based perspective-taking interactive fiction holds promise for reducing occupational stigma toward dirty work. By immersing participants in dynamic first-person scenarios, our intervention fosters empathy, deepens occupational understanding, and evokes resonance. Our findings further reveal that immersive experiences, a sense of professional achievement, and insights into the day-to-day lives of dirty workers are critical in mitigating stigma. Despite remaining challenges—such as the potential reinforcement of stereotypes and the need for multimodal engagement—this study provides empirical evidence that AI-powered perspective-taking can mitigate biases against dirty workers. Looking ahead, integrating richer multimodal content and exploring diverse contexts are expected to further expand the potential of LLM in advancing social equity.

\section*{Ethical Considerations}
Each study was reviewed and approved by the Institutional Review Board (IRB) at an accredited institution through a formal ethics review process. We adhered to rigorous procedures to safeguard the confidentiality, anonymity, and privacy of all participants. Personally identifiable information was collected solely for the purpose of distributing gratuities and was securely stored separately from the research data. Participants were assigned unique identifiers, and any additional information that could potentially compromise their anonymity was removed. Participation in the study was entirely voluntary. At the beginning of each session, participants were guided through an informed consent process, during which the study’s objectives and the intended use of the data were clearly explained. Participants were informed that if they experienced any discomfort while engaging with the LLM-based interactive fiction during the session, they could either skip any question or section they found uncomfortable or withdraw from the study at any time without penalty (i.e., they would still receive the gratuity). Participants were also assured that, should they choose to withdraw, all associated study data would be promptly deleted. In practice, no participants elected to withdraw.

\begin{acks}
We would like to thank our participants for their participation and for generously sharing their experiences. We also extend our gratitude to the anonymous reviewers for their constructive feedback, which greatly improved this work. We are also deeply grateful to all co-authors of this work. Their valuable contributions, encouragement, and unwavering support were essential in bringing this research to fruition.
\end{acks}

\bibliographystyle{ACM-Reference-Format}
\bibliography{references}

\appendix
\section{Appendix}
This appendix provides supplementary materials, methodological details, additional analyses, and conversation examples that extend our main findings. These materials offer deeper insights into our research process and data that could not be included in the primary manuscript due to space constraints.

\subsection{Evaluation and Refinement of the IF}
\label{appendix:1}
To further ensure that the IF we developed accurately simulated real-life situations of dirty work, we recruited professionals with at least three years of full-time work experience in the corresponding occupations through online platforms. These individuals served as evaluators rather than general participants. After completing the LLM-based first-person interactive fiction, they assessed the degree of similarity between the simulated scenarios and their actual work experiences using a 7-point Likert scale. The evaluation results are presented in \autoref{tab:occupations}. 
Their feedback then informed an iterative process of refining the prompt designs to enhance the realism and credibility of the simulations.

After completing the experience, we asked participants if there were any areas for improvement. They provided valuable feedback, such as one police officer noting, ``The scenario simulated by the bot mentions issuing a parking fine, but in China, this is the responsibility of traffic police, not general law enforcement officers.'' Based on this feedback, we implemented targeted improvements to the prompts, directly enhancing the accuracy and realism of the simulated scenarios.

\begin{table}[ht]
\centering
\small
\caption{Evaluation of the LLM-based Perspective-taking Interactive Fiction's Realism Compared to Actual Job Experiences}
\label{tab:occupations}
\begin{tabular}{@{}l p{1.5cm} p{2cm} p{2.5cm}@{}}
\toprule
\textbf{Occupation} & \textbf{Number of People} & \textbf{Average Years of Work} & \textbf{Average Similarity Rating} \\
\midrule
Janitor      & 4 & 4.5   & 5.75 \\
Police       & 4 & 7.25  & 6.25 \\
Firefighter  & 4 & 6.5   & 5.25 \\
Caregiver    & 4 & 4.25  & 5.75 \\
\bottomrule
\end{tabular}
\end{table}

\subsection{LLM Prompt Templates}
\label{appendix:prompt}

In the design of the entire prompt, we adhered to several key principles\cite{giray2023prompt,chen2023unleashing,marvin2023prompt}. First, we ensured task clarity by designing the entire task with clear objectives and a structured framework. Second, we prioritized simplicity and clarity, and included example inputs(few-shot learning) within the prompt. Furthermore, we made several adjustments to the prompt based on the output results. Here is an example of the prompt we used to construct the perspective-taking IF for police officers. The prompts for the other three occupations follow a similar structure, with the primary difference being the specific content:

\textbf{Objective:}

You will help the user experience a day in the life of a police officer, simulating the routine tasks and pressures that officers face. This simulation focuses on the mundane, stressful, and challenging aspects of police work, rather than high-profile cases. The user will interact with various scenarios and make decisions, gaining insight into the reactive and judgmental skills required in everyday police work.

\textbf{Structure and Process:}

Opening Guide:
You will begin by simulating a day in the life of a police officer. The tasks involved do not concern large-scale cases but highlight the everyday challenges of policing, such as patrolling, dealing with community issues, and managing minor conflicts. Small narrative elements can also be included to reflect the biases and misunderstandings that police officers, as part of a ``dirty work'' profession, might face. However, avoid including scenarios like traffic accidents or issuing parking tickets, as these typically fall under the purview of traffic police rather than regular officers. Additionally, you can incorporate moments of the officer's personal life, such as returning home after a shift only to be called back for an emergency, or being scolded by family members for being too busy with work and not spending enough time with them. Each scenario will offer several possible solutions. The user can choose from various options to understand the pressure and decision-making involved in police work.

Scenario Design and User Interaction:
After describing each scenario, you will present the user with several possible choices (behavioral options). Users may also suggest other choices they think a police officer might make. Each scenario and choice should reflect the pressures, trivialities, and interpersonal challenges inherent in police work. Allow users some freedom in their choices, but provide suggestions where necessary based on the context. You may offer hints to guide users in their decisions while allowing them to propose their own ideas. Each decision will influence the direction of the simulation, and you will advance the narrative based on the user's choices. 

Example:``You receive a report of a loud argument in the park. Upon arriving, you find two individuals arguing intensely, with emotions escalating. How will you handle the situation?''

Options:
(1) Attempt to calm both parties down and see if the conflict can be resolved. (2) Call for backup to assist in managing the situation. (3) Once they have calmed down, record the details and issue a report. (4) Users may also enter their own suggested response.

Handling Complex Situations:
When dealing with certain scenarios, prompt the user to balance emotional management with professional handling (e.g., when addressing sensitive issues such as domestic violence or missing persons). Encourage rational decision-making and provide professional advice on how officers should respond in these situations.

Scenario Feedback and Continuation:
After each user decision, provide real-time feedback to show the potential consequences of their choice. Advance the narrative based on their selection.
For example, if the user chooses ``request support,'' you can describe the arrival of backup officers and the collective handling of the situation.

Feedback on User Decisions:
Offer immediate feedback for each decision made by the user to help them understand the impact of their choices. For example: ``You chose to calm the bystanders, which helped stabilize the scene and prevent further escalation.''

Scenario Continuation:
Based on the user's choices, continue the scenario. Ensure that each situation not only provides a single outcome but leads the user into new challenges. For instance, if the user chooses ``call the police immediately'' during a community dispute, the next step could involve describing the officers’ further actions on the scene. Alternatively, if they choose to ``try to persuade the parties to calm down,'' describe the effects of this approach.

Realistic Police Tasks:
Scenarios should involve routine police tasks such as patrolling, managing traffic, dealing with neighborhood disputes, or handling minor cases encountered during regular patrols (e.g., theft, small conflicts, missing persons reports). Avoid overly dramatic situations, as the goal is to present a more authentic depiction of police work. Additionally, scenarios should not be confined to a full day's work; interludes, such as a police officer preparing for lunch after a busy morning shift only to be called back for an urgent task, could better highlight the unpredictability of the job and enhance the user's immersive experience.

Multiple Decision Paths:
For each scenario, offer at least three potential choices, ensuring that users experience different decision outcomes and consequences.

Providing Feedback and Summary:
After each decision, provide immediate feedback to help the user understand the consequences of their actions. After the simulation concludes, offer a global summary, reminding the user that the tasks they encountered are only a snapshot of a police officer’s daily responsibilities, and real-life situations may be even more complex and high-pressure.

Life Context Description:
In addition to work-related scenarios, consider adding a few that focus on the officer's personal life. For example, after returning home in the evening, the officer might be called back for a task and miss a family member’s birthday, or experience additional pressure from home. These situations will give the user a more holistic view of the challenges faced by police officers beyond their professional duties.

\textbf{Summary Reminder:}

At the end of the simulation, provide the user with a summary and reflection, emphasizing that the tasks experienced are just a part of the daily responsibilities of police officers. Remind them that real police work often involves more complex and unexpected situations.

``Your tasks today focused primarily on routine patrols, handling community issues, and managing minor conflicts. While these tasks may seem simple, they are often full of challenges and require officers to remain calm and efficient. In real-life policing, the complexity and pressure are even greater. We hope this experience has helped you better understand the difficulties of police work and has fostered greater empathy for the profession. Although you have experienced just a portion of a police officer's routine work today, real-life officers face far more. Every day brings uncertainty, from handling emergencies and interacting with the public to team coordination—all requiring swift decisions and sound judgment. We hope this simulation has allowed you to experience the daily struggles and challenges of the profession.''

\subsection{Custom-Designed Occupation Stigma Scale}
\label{appendix:scale}
Participants rated each item on a 7-point Likert scale (1 = Not at all familiar, 7 = Extremely familiar).

\textbf{Professional Knowledge}
\begin{itemize}
    \item To what extent are you familiar with the main tasks and responsibilities associated with this profession?
    \item To what extent do you understand the skills or qualifications required to pursue this profession?
    \item How familiar are you with the typical working hours and work environment of this profession?
\end{itemize}

\textbf{Occupational Stress and Risks}
\begin{itemize}
    \item How familiar are you with the general level of job stress experienced in this profession?
    \item To what extent do you understand the primary sources of occupational stress faced by professionals in this field?
    \item How familiar are you with the typical work situations that professionals in this field might encounter?
    \item To what extent are you aware of the potential risks associated with this profession?
    \item To what extent do you believe job-related stress impacts the quality of life of professionals in this field?
\end{itemize}

\textbf{Perceived Value of the Profession}
\begin{itemize}
    \item How would you evaluate the contribution of this profession to society?
    \item To what extent do you believe the absence of professionals in this field would affect the functioning of society?
\end{itemize}

\subsection{Supplementary Tables and Statistical Analyses}
\label{tables}
This section presents additional statistical analyses and supplementary tables that further support our main findings. These tables include detailed comparisons between occupation groups, comprehensive post-intervention measurements, and statistical results that were referenced but not fully displayed in the main text due to space constraints.

\begin{table*}[b]
\centering
\caption{Comparison of Empathy Traits Across Groups Using IRI and Beliefs About Empathy Scales. No significant differences were found in participants' baseline empathy levels across the four dirty work professions (Janitors, Firefighters, Police Officers, Caregivers), ensuring that subsequent results were not biased by initial empathy traits.}

\label{tab:IRI}
\begin{tabular}{@{}lcccccccccc@{}}
\toprule
 & \multicolumn{5}{c}{\textbf{IRI}} 
 & \multicolumn{5}{c}{\textbf{Beliefs about Empathy}} \\ 
\cmidrule(lr){2-6}\cmidrule(lr){7-11}
 & \textbf{n} & \textbf{Mean} & \textbf{S.D.} & \textbf{F} & \textbf{P} 
 & \textbf{n} & \textbf{Mean} & \textbf{S.D.} & \textbf{F} & \textbf{P} \\ 
\midrule
Janitors       
  & 25 & 3.567 & 0.184 
    & \multirow{4}{*}{1.121} & \multirow{4}{*}{0.345}
  & 25 & 4.555 & 0.795
    & \multirow{4}{*}{2.341} & \multirow{4}{*}{0.078} \\

Firefighters   
  & 25 & 3.538 & 0.167 
    & & 
  & 25 & 4.715 & 0.612
    & & \\

Police Officers 
  & 25 & 3.613 & 0.159
    & & 
  & 25 & 4.775 & 0.667
    & & \\

Caregivers     
  & 25 & 3.536 & 0.163 
    & & 
  & 25 & 4.335 & 0.447
    & & \\
\bottomrule
\end{tabular}
\end{table*}

\begin{table*}[ht]
\centering
\caption{Post-Experiment Scores for Empathy, Distress, and IOS Across Groups. Participants showed high levels of empathy, distress, and relational closeness with dirty workers across all groups, with no significant differences observed among the professions.}
\label{tab:empathy-distress-ios}
\begin{tabular}{@{}lccccc ccccc ccccc@{}}
\toprule
 & \multicolumn{5}{c}{\textbf{Empathy}} 
 & \multicolumn{5}{c}{\textbf{Distress}}
 & \multicolumn{5}{c}{\textbf{IOS}} \\
\cmidrule(lr){2-6}\cmidrule(lr){7-11}\cmidrule(lr){12-16}
 & \textbf{n} & \textbf{Mean} & \textbf{S.D.} & \textbf{F} & \textbf{P}
 & \textbf{n} & \textbf{Mean} & \textbf{S.D.} & \textbf{F} & \textbf{P}
 & \textbf{n} & \textbf{Mean} & \textbf{S.D.} & \textbf{F} & \textbf{P} \\
\midrule
\textbf{Janitors} 
  & 25 & 4.66 & 0.854 &  \multirow{4}{*}{0.209} &  \multirow{4}{*}{0.89}
  & 25 & 4.51 & 0.855 &  \multirow{4}{*}{1.634} &  \multirow{4}{*}{0.187}
  & 25 & 4.88 & 1.301  &  \multirow{4}{*}{0.326} &  \multirow{4}{*}{0.807} \\
\textbf{Firefighters}
  & 25 & 4.77 & 0.725   &      &    
  & 25 & 4.55 & 0.7  &      &    
  & 25 & 5.04 & 0.889   &      &     \\
\textbf{Police Officers}
  & 25 & 4.71 & 0.514 &      &    
  & 25 & 4.13 & 0.711 &      &    
  & 25 & 4.96 & 1.207   &      &     \\
\textbf{Caregivers}
  & 25 & 4.80 & 0.586  &      &    
  & 25 & 4.23 & 0.944 &      &    
  & 25 & 4.72 & 1.339   &      &     \\
\bottomrule
\end{tabular}
\end{table*}

\begin{table*}[ht]
\centering
\caption{Results of Caregivers. Significant improvements were observed in all measured areas.}
\label{tab:caregivers}
\begin{tabular}{@{}l l c c c c l c c c c@{}}
\toprule
\textbf{Items} & \textbf{Status} & \textbf{N} & \textbf{Mean} & \textbf{S.D.} & \textbf{Normality} & \textbf{Test Type} & \textbf{T} & \textbf{P} & \textbf{95\% CI} & \textbf{Cohen's |d|} \\
\midrule
\multirow{2}{*}{\textbf{Professional Knowledge}} 
  & Pre-test  & 25 & 2.213 & 0.552 & Y & Paired t-test & -5.986 & 0 & [-2.044, -0.996] & 1.588 \\
  & Post-test & 25 & 3.733 & 1.236 & Y &               &        &   &                &   \\
\multirow{2}{*}{\textbf{Occupational Stress and Risk}}
  & Pre-test  & 25 & 2.400 & 0.462 & Y & Paired t-test & -7.406 & 0 & [-1.770, -0.998] & 1.971 \\
  & Post-test & 25 & 3.784 & 0.879 & Y &               &        &   &                &   \\
\multirow{2}{*}{\textbf{Occupational Value Cognition} }
  & Pre-test  & 25 & 2.800 & 0.722 & Y & Paired t-test & -5.710 & 0 & [-1.307, -0.613] & 1.485 \\
  & Post-test & 25 & 3.760 & 0.561 & Y &               &        &   &                &   \\
\multirow{2}{*}{\textbf{Total}}
  & Pre-test  & 25 & 2.471 & 0.401 & Y & Paired t-test & -7.407 & 0 & [-1.647, -0.920] & 2.013 \\
  & Post-test & 25 & 3.760 & 0.812 & Y &               &        &   &                &   \\
\bottomrule
\end{tabular}
\end{table*}

\begin{table*}[ht]
\centering
\caption{Results of Police Officers. Significant improvements were observed in "Professional Knowledge," "Occupational Stress and Risk," and "Total," while "Occupational Value Cognition" showed no significant change.}
\label{tab:police_officers}
\begin{tabular}{@{}l l c c c c l c c c c@{}}
\toprule
\textbf{Items} & \textbf{Status} & \textbf{N} & \textbf{Mean} & \textbf{S.D.} & \textbf{Normality} & \textbf{Test Type} & \textbf{T} & \textbf{P} & \textbf{95\% CI} & \textbf{Cohen's |d|} \\
\midrule
\textbf{Professional Knowledge} 
  & Pre-test  & 25 & 3.053 & 1.053 & Y & Paired t-test & -4.881 & 0 & [-1.878, -0.762] & 1.418 \\
  & Post-test & 25 & 4.373 & 0.790 & Y &               &        &   &                &   \\
\textbf{Occupational Stress and Risk} 
  & Pre-test  & 25 & 3.984 & 0.911 & Y & Paired t-test & -3.856 & 0.001 & [-1.486, -0.450] & 1.096 \\
  & Post-test & 25 & 4.952 & 0.855 & Y &               &        &   &                &   \\
\textbf{Occupational Value Cognition} 
  & Pre-test  & 25 & 5.000 & 1.283 & Y & Paired t-test & -0.676 & 0.505 & [-0.972, 0.492] & 0.180 \\
  & Post-test & 25 & 5.240 & 1.378 & Y &               &        &   &                &   \\
\textbf{Total} 
  & Pre-test  & 25 & 3.848 & 0.673 & Y & Paired t-test & -6.364 & 0 & [-1.334, -0.681] & 1.543 \\
  & Post-test & 25 & 4.855 & 0.632 & Y &               &        &   &                &   \\
\bottomrule
\end{tabular}
\end{table*}

\begin{table*}[ht]
\centering
\caption{Results of Janitors. Significant improvements were observed in all areas.}
\label{tab:janitors}
\begin{tabular}{@{}l l c c c c l c c c c@{}}
\toprule
\textbf{Items} & \textbf{Status} & \textbf{N} & \textbf{Mean} & \textbf{S.D.} & \textbf{Normality} & \textbf{Test Type} & \textbf{T} & \textbf{P} & \textbf{95\% CI} & \textbf{Cohen's |d|} \\
\midrule
\textbf{Professional Knowledge} 
  & Pre-test  & 25 & 2.690 & 1.090 & Y & Paired t-test & -5.423 & 0 & [-2.301, -1.032] & 1.846 \\
  & Post-test & 25 & 4.360 & 0.670 & Y &               &        &   &                &   \\
\textbf{Occupational Stress and Risk} 
  & Pre-test  & 25 & 2.830 & 1.110 & Y & Paired t-test & -5.074 & 0 & [-2.228, -0.940] & 1.636 \\
  & Post-test & 25 & 4.420 & 0.810 & Y &               &        &   &                &   \\
\textbf{Occupational Value Cognition} 
  & Pre-test  & 25 & 2.360 & 0.760 & Y & Paired t-test & -6.255 & 0 & [-1.995, -1.005] & 1.442 \\
  & Post-test & 25 & 3.860 & 1.260 & Y &               &        &   &                &   \\
\textbf{Total} 
  & Pre-test  & 25 & 2.520 & 0.780 & Y & Paired t-test & -6.731 & 0 & [-2.252, -1.195] & 2.276 \\
  & Post-test & 25 & 4.250 & 0.740 & Y &               &        &   &                &   \\
\bottomrule
\end{tabular}
\end{table*}

\begin{table*}[ht]
\centering
\caption{Results of Firefighters. Significant improvements were observed in "Professional Knowledge," "Occupational Stress and Risk," and "Total," while "Occupational Value Cognition" showed no significant change.}
\label{tab:firefighters}
\begin{tabular}{@{}l l c c c c l c c c c@{}}
\toprule
\textbf{Items} & \textbf{Status} & \textbf{N} & \textbf{Mean} & \textbf{S.D.} & \textbf{Normality} & \textbf{Test Type} & \textbf{T} & \textbf{P} & \textbf{95\% CI} & \textbf{Cohen's |d|} \\
\midrule
\textbf{Professional Knowledge} 
  & Pre-test  & 25 & 2.560 & 0.699 & Y & Paired t-test & -6.554 & 0 & [-2.139, -1.114] & 2.123 \\
  & Post-test & 25 & 4.187 & 0.828 & Y &               &        &   &                &   \\
\textbf{Occupational Stress and Risk} 
  & Pre-test  & 25 & 3.944 & 0.830 & Y & Paired t-test & -2.448 & 0.022 & [-0.944, -0.080] & 0.560 \\
  & Post-test & 25 & 4.456 & 0.991 & Y &               &        &   &                &   \\
\textbf{Occupational Value Cognition} 
  & Pre-test  & 25 & 4.200 & 1.109 & Y & Paired t-test & -1.731 & 0.096 & [-1.271, 0.111] & 0.539 \\
  & Post-test & 25 & 4.780 & 1.042 & Y &               &        &   &                &   \\
\textbf{Total} 
  & Pre-test  & 25 & 3.732 & 0.660 & Y & Paired t-test & -3.953 & 0.001 & [-1.129, -0.354] & 1.057 \\
  & Post-test & 25 & 4.474 & 0.740 & Y &               &        &   &                &   \\
\bottomrule
\end{tabular}
\end{table*}

\begin{table*}[ht]
\centering
\caption{Results of Multiple Comparisons with Bonferroni Correction. No statistically significant differences were found between occupational groups across all measures (Beliefs about Empathy, IOS, IRI, Empathy, and Distress), as all adjusted p-values were nonsignificant.}
\label{tab:bonferroni_updated}
\begin{tabular}{@{}l l l c c c@{}}
\toprule
\textbf{Item} & \textbf{Work Type 1} & \textbf{Work Type 2} & \textbf{P} & \textbf{95\% CI Lower} & \textbf{95\% CI Upper} \\
\midrule
\multirow{6}{*}{\textbf{Distress}} 
    & Janitors & Firefighters & 0.600 & -0.236 & 0.996 \\
    & Janitors & Police Officers & 1.000 & -0.656 & 0.576 \\
    & Janitors & Caregivers & 1.000 & -0.336 & 0.896 \\
    & Firefighters & Police Officers & 0.417 & -1.036 & 0.196 \\
    & Police Officers & Caregivers & 0.991 & -0.296 & 0.936 \\
    & Caregivers & Firefighters & 1.000 & -0.516 & 0.716 \\
\tabularnewline
\multirow{6}{*}{\textbf{Beliefs about Empathy}} 
    & Janitors & Firefighters & 1.000 & -0.270 & 0.710 \\
    & Janitors & Police Officers & 1.000 & -0.650 & 0.330 \\
    & Janitors & Caregivers & 1.000 & -0.710 & 0.270 \\
    & Firefighters & Police Officers & 0.235 & -0.870 & 0.110 \\
    & Police Officers & Caregivers & 1.000 & -0.550 & 0.430 \\
    & Caregivers & Firefighters & 0.104 & -0.050 & 0.930 \\
\tabularnewline
\multirow{6}{*}{\textbf{IOS}} 
    & Janitors & Firefighters & 1.000 & -0.990 & 0.830 \\
    & Janitors & Police Officers & 1.000 & -1.070 & 0.750 \\
    & Janitors & Caregivers & 1.000 & -0.750 & 1.070 \\
    & Firefighters & Police Officers & 1.000 & -0.990 & 0.830 \\
    & Police Officers & Caregivers & 1.000 & -0.590 & 1.230 \\
    & Caregivers & Firefighters & 1.000 & -1.150 & 0.670 \\
\tabularnewline
\multirow{6}{*}{\textbf{IRI}} 
    & Janitors & Firefighters & 1.000 & -0.174 & 0.083 \\
    & Janitors & Police Officers & 1.000 & -0.099 & 0.157 \\
    & Janitors & Caregivers & 1.000 & -0.097 & 0.159 \\
    & Firefighters & Police Officers & 0.725 & -0.054 & 0.203 \\
    & Police Officers & Caregivers & 1.000 & -0.127 & 0.130 \\
    & Caregivers & Firefighters & 0.673 & -0.205 & 0.052 \\
\tabularnewline
\multirow{6}{*}{\textbf{Empathy}} 
    & Janitors & Firefighters & 1.000 & -0.570 & 0.470 \\
    & Janitors & Police Officers & 1.000 & -0.630 & 0.410 \\
    & Janitors & Caregivers & 1.000 & -0.660 & 0.380 \\
    & Firefighters & Police Officers & 1.000 & -0.580 & 0.460 \\
    & Police Officers & Caregivers & 1.000 & -0.550 & 0.490 \\
    & Caregivers & Firefighters & 1.000 & -0.430 & 0.610 \\
\bottomrule
\end{tabular}
\end{table*}

\begin{table*}[ht]
\centering
\caption{Components of Meaning and Example Questions. This table illustrates the key components of meaning derived from the perspective-taking simulation experiences and provides example questions used to explore participants' thoughts and feelings. The components include \textit{Emotional Resonance, Social Bias and Occupational Stigma, and Reflection on the LLM-based perspective-taking interactive fictio}n. The listed questions aim to investigate participants' understanding of the tasks, challenges, and biases associated with janitors, firefighters, police officers, and caregivers, as well as their reflections on the effectiveness and limitations of the simulation approach.}
\label{tab:example}
\begin{tabular}{m{4cm} m{10cm}}
\toprule
\centering \textbf{Components of Meaning} 
& \centering \textbf{Example Questions} \tabularnewline
\midrule
\centering \textbf{Emotional Resonance}
& 
\begin{itemize}
    \item After experiencing the virtual scenarios of these professions, did you gain a deeper understanding of the specific tasks of janitors/firefighters/police officers/caregivers?
    \item During the experience, did you feel the stress, fatigue, or sense of accomplishment commonly associated with these professions? Please describe your feelings.
    \item Did you develop any sympathy or empathy for professionals in these fields during the experience? If so, what factors contributed to these emotions?
    \item Did the simulation experience change your perception and attitude toward these professions? If so, what do you believe caused this shift in your views?
\end{itemize}
\tabularnewline
\centering \textbf{Social Bias and Occupational Stigma}
& 
\begin{itemize}
    \item Have you ever held negative views about this profession? If so, what factors contributed to these views?
    \item After experiencing these professions, did your negative perceptions of them change?
    \item Do you believe that the social bias these professions face is unrelated to the actual work content or social role of these occupations? Why?
    \item Do you think professionals in these fields deserve more respect and support? Why?
\end{itemize}
\tabularnewline
\centering \textbf{Reflection on the LLM-based perspective-taking interactive fiction}
& 
\begin{itemize}
    \item How much do you think this LLM-based perspective-taking interactive fiction helped reduce your bias and improve your empathy?
    \item Do you consider this LLM-based perspective-taking interactive fiction an effective method for changing biases? Why?
    \item What do you think are the advantages of this approach?
    \item Do you think there are any limitations or challenges associated with this experience? Why?
    \item In the future, would you be willing to try more similar simulation experiences? Do you believe this virtual experience approach could be scaled to change the cognition and biases of other social groups?
\end{itemize}
\tabularnewline
\bottomrule
\end{tabular}
\label{tab:components}
\end{table*}

\clearpage
\subsection{Interactive Fiction Conversations}
This section presents an exemplar conversation from our interactive fiction experiments, specifically showcasing P12's dialogue during the Caregiver occupation simulation. The following exchange illustrates the typical flow, depth, and nature of perspective-taking experiences facilitated through our LLM-based intervention.
\label{dialogue}
\tiny

LLM: \quoted{You are about to simulate a day in the life of a caregiver. Type Start when you’re ready, and we’ll begin this interactive experience!}\\

User: \quoted{Start}\\

LLM: \quoted{Great! Let’s begin simulating a caregiver’s day. Morning: 6:30 AM You arrive early at your employer’s home — a family of four: two parents and two children (ages 3 and 7). Your first task is to prepare breakfast for the family. However, just as you start, the 3-year-old runs out of the bedroom crying, saying he’s hungry. Scenario 1: The Hungry, Crying Child You are chopping vegetables for breakfast while the child’s crying gets louder. What would you do? Options: 1. Stop preparing breakfast, and give the child some snacks to calm him down. 2. Tell the child to wait five more minutes and continue cooking. 3. Pick up the child, comfort him, and try to continue preparing breakfast at the same time. Please tell me your choice (enter 1/2/3), or describe your own approach!}\\

User: \quoted{Do 1 first, then 3.}\\

LLM: \quoted{Understood! You chose to first give the child some snacks to calm him (Option 1), and then pick him up, comfort him, and try to continue preparing breakfast (Option 3). Feedback: You give the child some small biscuits, and he gradually stops crying as he eats. Then, you pick him up, gently comforting him and telling him breakfast will be ready soon. The child seems a bit more reassured. However, holding him makes it a little harder for you to prepare breakfast, slightly slowing you down. Outcome: • The child’s emotions were soothed, and he stopped crying after eating some snacks. • Breakfast was slightly delayed, but you still managed to chop the vegetables and continue working. • When the family woke up, they noticed the child was calm and thanked you.}\\

LLM: \quoted{Scenario 2: Challenges at the Breakfast Table At 7:30 AM, you finally finish making breakfast (fried eggs, toast, and fruit). The family sits down to eat, but the 7-year-old complains about disliking the fried eggs and refuses to eat. Meanwhile, the 3-year-old starts grabbing fruit with his hands, spilling juice all over the table. Your choices: 1. Try to talk with the 7-year-old, encourage him to eat, while helping the 3-year-old eat fruit with a spoon. 2. Let the 7-year-old decide whether to eat or not, and focus on helping the 3-year-old keep the table clean. 3. Clean up the juice first, and then talk to the 7-year-old about breakfast. Please tell me your choice (enter 1/2/3), or describe your own approach!}\\

User: \quoted{2}\\

LLM: \quoted{You chose Option 2: let the 7-year-old decide whether to eat, and focus on helping the 3-year-old keep the table clean. Feedback: You decided not to intervene too much with the 7-year-old and respected his choice. Meanwhile, you quickly grabbed tissues to clean up the spilled juice and encouraged the 3-year-old to use a spoon to eat fruit. Outcome: • The 7-year-old skipped breakfast but felt relieved that you didn’t force him and started playing with toys instead. • The 3-year-old attempted to eat with a spoon under your guidance but still got his clothes dirty. • The table was tidied, though you’d need to wash the child’s clothes later.}\\

LLM: \quoted{Scenario 3: Morning Accident At 10:00 AM, while you’re cleaning the living room floor and watching the kids play, the 3-year-old trips and scrapes his knee, crying loudly. The 7-year-old shouts for you to come quickly. Your choices: 1. Immediately drop the cleaning tools, check the child’s injury, and comfort him. 2. Ask the 7-year-old to help calm his brother while you quickly finish cleaning before going over. 3. Comfort the 3-year-old verbally while continuing to clean, telling him it’s a small injury and not to worry. Please tell me your choice (enter 1/2/3), or describe your own approach!}\\

User: \quoted{1}\\

LLM: \quoted{You chose Option 1: immediately drop your cleaning tools, check the injury, and comfort the child. Feedback: You quickly set aside your cleaning tools, rushed to the 3-year-old, crouched down to check his scraped knee. It wasn’t serious, but a bit red and swollen. You comforted him gently, saying, “It’s okay, you’ll feel better soon.” Then you fetched the first-aid kit, carefully cleaned the wound, and applied a band-aid. The 7-year-old saw you tending to his brother attentively and calmed down as well. Outcome: • The 3-year-old stopped crying after your careful care and went back to playing. • The 7-year-old admired your calm response and even helped you tidy up his brother’s toys. • Your cleaning task was delayed, but the children’s safety and emotions were well handled.}\\

LLM: \quoted{Scenario 4: Lunchtime Update As noon approaches, you begin preparing lunch — simple pasta and salad as requested by the employer. Just then, the 7-year-old runs over and says, “Mom said we’re not eating at home today. We’re going out.” You’re a bit confused, as the employer hadn’t informed you of this change. Your choices: 1. Pause cooking and text the employer to confirm whether to continue making lunch. 2. Continue cooking as planned and wait for the employer’s return to decide. 3. Prepare a few simple backup snacks just in case the children get hungry, while waiting for further instructions. Please tell me your choice (enter 1/2/3), or describe your own approach!}\\

User: \quoted{1}\\

LLM: \quoted{You chose Option 1: pause cooking and text the employer to confirm. Feedback: You paused your cooking, took out your phone, and messaged the employer to ask whether to continue preparing lunch. A few minutes later, they replied, “Sorry, I forgot to tell you — we’ll be eating out today. No need to cook, thank you!” Though this change wasted some of your time, your timely communication prevented further unnecessary work. The kids, excited about eating out, began getting ready. Outcome: • You avoided wasting food and effort. • The employer appreciated your communication and flexibility. • You gained a bit of extra time to tidy up the kitchen or take a short break.}\\

LLM: \quoted{Scenario 5: Afternoon Request At 3:00 PM, after the family returns home from lunch, you finally have a breather. Suddenly, your employer says, “We’re having friends over tonight. Could you help prepare dinner and tidy the living room?” Your original shift ends at 6:00 PM, meaning you might need to work overtime. Your choices: 1. Agree to the request and start preparing dinner and cleaning as needed. 2. Politely explain that you have other plans and cannot work overtime. 3. Ask if it’s possible to adjust the workload and see if you could help partially. Please tell me your choice (enter 1/2/3), or describe your own approach!}\\

User: \quoted{I would ask what time I’m expected to finish and whether there’s additional overtime pay.}\\

LLM: \quoted{You chose to inquire about the estimated end time and the possibility of overtime pay — a very reasonable and professional approach! Feedback: You politely asked, “Could you let me know what time I’d be expected to finish tonight? Also, will there be overtime pay if I stay later?” The employer hesitated a bit but expressed understanding. “We hope you could stay until around 8:00 PM to help with dinner and cleaning. Overtime will be paid hourly, the same as your regular rate.” You agreed, clarified your working hours and compensation, and began preparing for the evening tasks. Outcome: • You clearly communicated your needs and gained the employer’s understanding. • With confirmed overtime pay, your additional work was fairly compensated. • The employer appreciated your communication and promised to notify you in advance for future overtime needs.}\\

LLM: \quoted{Scenario 6: Kitchen Disruption While preparing dinner, one of the employer’s friends arrives early with their 5-year-old, who runs into the kitchen and starts commenting on the food, saying, “This is yucky, I don’t like this!” potentially disrupting your work. Your choices: 1. Politely tell the child that the kitchen is dangerous and ask them to go play in the living room. 2. Try to communicate with the child, ask what they like to eat, and adjust the menu if possible. 3. Ignore the child’s interruptions and continue cooking. Please tell me your choice (enter 1/2/3), or describe your own approach!}\\

User: \quoted{Do 1 first, then 2.}\\

LLM: \quoted{You chose to first politely ask the child to leave the kitchen (Option 1), then talk to them about their food preferences (Option 2). Feedback: You crouched down and gently told the child, “Sweetheart, the kitchen has dangerous things like knives and hot pans — it’s not safe for you here. Can you go play in the living room with the others?” Although a little reluctant, the child left after your patient persuasion. Later, you approached him and asked, “I heard you don’t like some of the food. What’s your favorite?” The child happily replied, “I like fried chicken and fries!” You explained that those weren’t on the menu tonight, but you’d add some simple items, like boiled eggs or mashed potatoes, to suit his taste as much as possible. Outcome: • You successfully removed the child from the kitchen, avoiding potential disruptions and safety risks. • Your communication made the child feel respected, improving his mood. • The employer praised your flexibility and communication skills after hearing how you handled it.}\\

LLM: \quoted{Scenario 7: Post-Dinner Cleanup After dinner, the living and dining areas are a mess. You start cleaning, but it’s already close to 8:00 PM, your agreed finishing time. The employer doesn’t mention your shift ending and seems to expect you to keep cleaning. Your choices: 1. Leave on time as agreed, politely reminding the employer that your work and overtime are completed. 2. Continue cleaning until everything is done, even if unpaid, to avoid awkwardness. 3. Communicate that you’re willing to help clean but hope for extra pay for the additional time. Please tell me your choice (enter 1/2/3), or describe your own approach!}\\

User: \quoted{3}\\

LLM: \quoted{You chose Option 3: communicate that you’re willing to help but would like additional pay. Feedback: You approached the employer and politely said, “Today’s work has already gone beyond the scheduled time. I’m happy to continue helping with the cleanup, but I hope this extra work can be compensated as overtime. Would that be alright?” The employer, though a little embarrassed, agreed. “Thank you so much for your help — we’ll pay for the extra time.” You continued cleaning, spending around 40 minutes washing dishes, wiping the table, and taking out the trash. Outcome: • You showed professionalism and flexibility while handling the cleanup well. • The employer appreciated your work and promised to add overtime pay. • Your direct but polite communication protected your rights while maintaining a good working relationship.}\\

LLM: \quoted{Scenario 8: Family Pressure After Work After a long, busy day, you finally get home after 9:00 PM. However, your partner and child seem upset about your late return. Your partner complains, “You’re always taking care of other people’s families — what about your own? Our kid barely remembers what you look like!” Your child says, “Mommy, why don’t you sleep with me again?” Your choices: 1. Try to communicate, explain your work situation, and promise to better balance time with them. 2. Stay silent and accept their frustration, feeling explanations won’t help. 3. Respond emotionally, arguing that you work hard for the family’s finances. Please tell me your choice (enter 1/2/3), or describe your own approach!}\\

User: \quoted{Option 1, but without making promises.}\\

LLM: \quoted{You chose Option 1: attempt to communicate and explain your work situation, but without making promises. Feedback: You calmly told your partner and child, “I understand you both want me to spend more time at home, but my job requires a lot of time and effort — especially today with overtime, I really couldn’t come back earlier. I truly care about our family, and I’m working hard so we can have a better life. I hope you can be a little more understanding.” Your tone was gentle, and while you didn’t make promises you couldn’t keep, you emphasized your commitment and care. Your partner, though still somewhat displeased, didn’t say more, and your child quieted down a little. Outcome: • Your calm communication eased family tensions. • You avoided making promises you couldn’t fulfill, protecting your emotions and relationships. • Balancing work and family remains a challenge, but you helped your loved ones understand you’re trying your best.}\\

LLM: \quoted{Summary \& Reflection: Today, you experienced the busy and complex work of a caregiver — from early morning childcare, lunchtime changes, overtime requests, to family conflicts at home. Key Takeaways: • Handling Emergencies: You responded calmly and flexibly when faced with falls and kitchen disruptions, ensuring safety and efficiency. • Employer Communication: You maintained professionalism by respecting employer needs while firmly upholding your own rights. • Family Pressure: Family expectations are a hidden challenge in caregiving jobs. Through communication, you eased conflicts, while acknowledging the need for greater support at home. Final Note: Today’s tasks revolved around family caregiving and daily chores — seemingly simple but full of challenges, requiring constant role-switching and effective communication. Caregiver work is a form of “dirty work” — often undervalued but crucial to family and societal functioning. Hopefully, through this simulation, you’ve gained deeper insight into the complexity and difficulty of caregiving work.}\\

\end{document}